\documentclass[prl,twocolumn]{revtex4}
\usepackage{graphicx}

\usepackage{epstopdf} \DeclareGraphicsExtensions{.eps, .pdf}

\begin{document}

\def\K{{\bf{K}}}
\def\Q{{\bf{Q}}}
\def\X{{\bf{X}}}
\def\Gbar{\bar{G}}
\def\tk{\tilde{\bf{k}}}
\def\k{{\bf{k}}}
\def\q{{\bf{q}}}
\def\x{{\bf{x}}}
\def\y{{\bf{y}}}

\title{Bond excitations in the  pseudogap phase of the 
Hubbard Model}
\author{Alexandru Macridin and  M.\ Jarrell}
\address{
University of Cincinnati, Cincinnati, Ohio, 45221, USA}

\date{\today}

\begin{abstract}

Using the dynamical cluster approximation, we calculate the correlation 
functions associated with the nearest neighbor bond operator which measure 
the z component of the spin exchange in the two-dimensional 
Hubbard model with $U$ equal to the bandwidth.  We find that in the 
pseudogap region, the local bond susceptibility diverges at $T=0$.  This 
shows the existence of degenerate bond spin excitation and implies 
quantum criticality and bond order formation when long range correlations 
are considered.  The strong correlation between  excitations on parallel 
neighboring bonds suggests bond singlet dimerization. The suppression of  
divergence for $n< \approx 0.78$ implies that tor these model parameters
this is quantum critical point which separates the unconventional pseudogap 
region characterized by bond order from a conventional Fermi liquid.

\end{abstract}

\maketitle


{\em{Introduction.}} The low doping pseudogap (PG) region of the cuprates 
has remained an issue of great discussion and controversy, with experimental 
data showing  anomalous behavior such as the suppression of spin excitations 
in the susceptibility, a PG in the single-particle spectra, and patterns in 
the STM spectra, among others\cite{PGreview,JCDavis}.  Different 
investigators argued  that the PG  is related with  the settlement of 
order~\cite{Chakravarty, Varma, Kivelson, Voijta, Sachdev}, where  the optimal 
doping is  in the proximity of the quantum critical point (QCP) associated 
with this order~\cite{Voijta}.  Previously\cite{macridin:cc} we investigated 
the staggered flux order in the PG region of single-band Hubbard model, as 
proposed by Chakravarty {\em et al}\cite{Chakravarty}. Despite the clear 
evidence of the PG signature in both single particle DOS and two particle 
magnetic spectrum, similar to experimental data in cuprates, we found no 
evidence of  staggered flux order.  

Here we investigate a different kind of order  associated with  
spin bond  correlations. Spin bond order states were proposed 
to take place in the PG region~\cite{Read, Sachdev, Sachdev_Park, Senthil}.
These bond orders require the investigation of four-particle susceptibilities,
which is presently very difficult to calculate with our method. However, while 
our method does not allow an exhaustive investigation of bond order states 
we find compelling evidence that in the PG region the bond magnetic degrees 
of freedom should order.

Investigating the local  bond excitation  susceptibility with the
dynamical cluster approximation (DCA)~\cite{hettler:dca,maier:rev}, we 
find evidence of  quantum criticality in the 2D Hubbard model. We consider  
the Coulomb interaction  $U$ to be equal  to the bandwidth  $W=8t$.  The 
DCA is a cluster mean-field theory which maps the original lattice model onto 
a periodic cluster of size $N_c=L_c^2$ embedded in a self-consistent host. 
Spatial correlations up to a range $L_c$ are treated explicitly, while those 
at longer length scales are described at the mean-field level. However the 
correlations in time, essential for local criticality,  are treated explicitly for all 
cluster sizes.  We measure the fluctuations associated with the nearest neighbor  
bond operator which measure the z-component of spin exchange on the bond. We 
find that there are degenerate bond spin  excitations  in the doping range 
$0\%- \approx 22\%$ corresponding to the PG region, which results in a divergent 
local bond susceptibility at $T=0$.  This divergence is caused by ordering in  
imaginary time rather than the more familiar ordering in  space, and  associated 
with the settlement of long range order at a  general phase transition. Nevertheless,  
in the limit $N_c \rightarrow \infty$  one should expect that long range bond 
correlations  will quench the entropy and a  transition to a state with long range 
order will take place~\cite{local_moment},  unless  a stronger instability such as 
d-wave pairing occurs first. The DCA method,  which does not allow spatial ordering 
on  distances larger than the cluster size, will fail to capture this transition when 
small clusters are considered. However at temperatures larger than the ordering 
temperature the physics would be determined predominantly  by the local quantum 
fluctuations described with DCA.  The divergent behavior of bond susceptibility   is 
suppressed for doping $> 22\%$ implying that for these model parameters, $22\%$ 
doping  is a QCP which separates the unconventional pseudogap region characterized 
by bond order from a conventional Fermi liquid.   We also find a strong correlation 
between  excitations on parallel neighboring bonds, which suggests that the pseudogap 
region is characterized by bond singlet dimerization.

{\em{Formalism.}}  To solve the  cluster problem we use the Hirsch-Fye 
quantum Monte Carlo (QMC) method\cite{j_hirsch_86a} which is based on a 
discrete path integral approximation with time step $\Delta\tau$.  
Hirsch-Hubbard-Stratonovich (HHS) fields are introduced to decouple the 
interaction\cite{j_hirsch_83}
\begin{equation}
\exp\left(
-\Delta\tau U n_\uparrow n_\downarrow 
+ \Delta\tau U (n_\uparrow + n_\downarrow)/2
\right)
= 
\frac12 {\rm{Tr}}_\sigma e^{2\alpha \sigma (n_\uparrow - n_\downarrow)} \,.
\end{equation}
where an Ising HHS decoupling field $\sigma=\pm 1$ is introduced at each
spin-time location on the cluster.  This  transforms the problem of interacting 
electrons to one of non-interacting particles coupled to time-dependent fields. 
The fermionic fields are then integrated out, and the integrals over the HHS 
decoupling fields are performed with a Monte Carlo algorithm.

All measurable quantities are completely determined by the HHS fields and 
the host Green's function.  The HHS fields contain all information about 
correlations (spin, pair and charge) in space  and time.  Moreover, Hirsch has 
shown that spin correlations may be directly rewritten in terms of the HHS 
decoupling fields\cite{j_hirsch_83,j_hirsch_86b}.   One may interpret the Ising 
fields as representing the fermion spin 
variables
\begin{equation}
\label{eq:hhssz}
\left( n(i,\tau)_{\uparrow}- n(i,\tau)_{\downarrow}\right) \to
\left(1-e^{-\Delta\tau U}\right)^{-1/2} \sigma(i,\tau)\,.
\end{equation}
Note that this is an exact mapping, so that all correlation functions of 
the HHS fields are, up to a proportionality constant, equivalent to the 
corresponding correlation functions of
$S^z(i,\tau)=\frac{1}{2}(n(i,\tau)_{\uparrow}- n(i,\tau)_{\downarrow})$.

\begin{figure}[t]
\begin{center}
\includegraphics*[height=2in, width=3.3in]{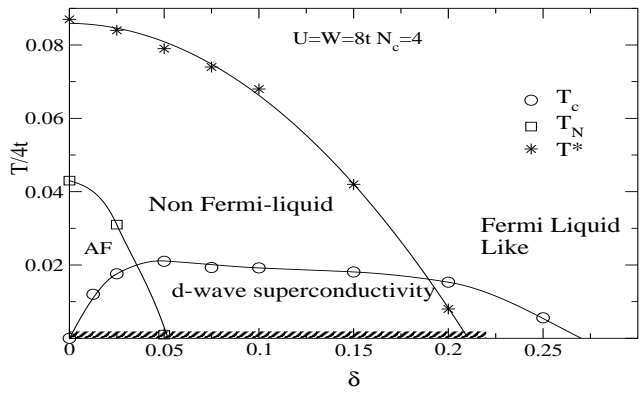}
\caption{Phase diagram of the Hubbard model for N$_c$=4 and $U=W=8t$. 
$T_c$, $T_N$ and $T^*$ are the superconducting,  antiferromagnetic and 
pseudogap temperatures from \cite{mark_phase}.  In the  marked region, 
$0<\delta<0.22$ doping, which corresponds to the PG regime, a divergent
local bond susceptibility is found. }
\label{fig:phasediagram}
\end{center}
\end{figure}

In the DCA the single-particle and two-particle lattice response functions 
are calculated with the Dyson equation using the irreducible cluster 
self-energy and vertices, respectively. Unlike experiments, where  in order to 
search for quantum critcality, a magnetic field is applied to suppress 
the superconductivity, in DCA the superconductivity can be suppressed
by imposing a normal state host.   We then look for divergent susceptibilities
in the normal state indicating phase boundaries.   Lattice  two-particle susceptibilities 
indicate transitions to antiferromagnetic (AF) and d-wave superconducting 
states at finite doping according to the phase diagram shown in 
Fig.~\ref{fig:phasediagram}.  However ordering associated with more complex 
operators, such as valence bond singlets\cite{Sachdev},  is far more difficult
to detect with the DCA, since it involves  complex equations and up to eight-leg 
irreducible interaction vertices.  More feasible calculations involve the corresponding 
cluster susceptibilities, since they can be obtained directly in the QMC process.  
However, these cluster susceptibilities are finite size quantities and can diverge 
only at zero temperature (i.e., infinite imaginary time when ordering in time 
occurs).

To study bond correlations, we define the bond $"ij"$ operator at time $\tau $ as
\begin{eqnarray}
\label{eq:Bi}
B(i,j;\tau) =  \sigma(i,\tau)\sigma(j,\tau)  \propto S^z(i,\tau) S^z(j,\tau)
\end{eqnarray}
\noindent where $"i"$  and $"j"$ label the position in the cluster.   For simplicity, 
we also denote with $B_{nn}$  ($B_{nnn}$) the bond operator when $"i"$  and 
$"j"$ are (next) nearest neighbor sites. 

 In the next section we present results for the correlation functions:
\begin{eqnarray}
\label{eq:chi0}
\chi_{0}(T) =&  \int d \tau \langle \delta B(i; i+ \hat x,\tau) \delta B(i; i+\hat x,0) \rangle \\
\label{eq:chipp}
\chi_{\perp}(T) = & \int d \tau  \langle \delta B(i; i+ \hat x,\tau) \delta B(i; i+\hat y,0) \rangle \\
\label{eq:chipa}
\chi_{\parallel}(T) =& \int d \tau  \langle \delta B(i; i+ \hat x,\tau) \delta B(i+\hat y; i+\hat x + \hat y,0) \rangle \\  \nonumber
\end{eqnarray}
\noindent where 
\begin{equation}
 \delta B(i; i+ \hat x,\tau) =B(i; i+ \hat x,\tau)-\langle B_{nn} \rangle~.
\label{eq:deltab}
\end{equation}
 
\noindent  We also measure $\chi_s$, the susceptibility associated with 
the operator $M$
\begin{eqnarray}
M (\tau)= &\frac{1}{N_c} \sum_i (B(i,i+\hat x;\tau)+B(i,i+\hat y;\tau)) \\
\chi_{s}(T) = & \int d \tau  \langle \delta
\langle M(\tau)\delta M(0) \rangle~. 
\label{eq:chis}
\end{eqnarray}
\noindent These correlation functions  describe the response of the system
to an  external field which couples with the bond operator $B_{nn}$.
The  field acts to modify the z-component of the nearest neighbor exchange 
interaction. Depending on its sign it decreases or increases the energy 
of an AF bond and has an opposite effect on a FM bond. 
$\chi_{0}$  describes the local bond response while  $\chi_{\perp}$  and 
$\chi_{\parallel}$ the correlation between nearest neighbor bonds. 
$\chi_s$ is a cluster quantity incorporating spatial correlations 
within the cluster between bonds.  

\begin{figure}[t]
\begin{center}
\includegraphics*[width=3.3in]{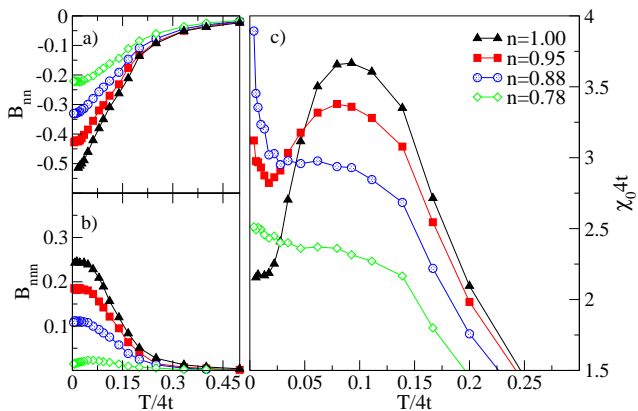}
\caption{(color online) a) ( b)) Nearest  (next-nearest) neighbor  bond expectation value $B_{nn}$ 
($B_{nnn}$) versus T for different  fillings $n$.   Short range AF order is present in the 
system.  c) Local bond susceptibility $\chi_0$ versus T. 
  $\chi_0$ show a divergent behavior when $T \rightarrow 0$ 
in the pseudogap region indicating critical behavior.}
\label{fig:chiaf}
\end{center}
\end{figure}

{\em{Results.}} 
We first present calculations on a $2 \times 2$ cluster, the smallest cluster 
capable of reproducing the generic features of the cuprate phase diagram. 
In the doping region relevant for high T$_c$ cuprates, the Hubbard model 
shows evidence of short range AF correlations. The expectation value of 
(next) nearest neighbor bond operator $B_{nn}$ ($B_{nnn}$) is  negative 
(positive) and increases with lowering  temperature, as one expects for a 
system with short range AF order.  The short range AF order is stronger 
at smaller doping. $B_{nn}$  and  $B_{nnn}$ versus temperature at 
different fillings are shown for a $N_c=4$ cluster in Fig.~\ref{fig:chiaf} -a) 
and, respectively, -b).

In the electron density range $1>n>\sim 0.78$,  the temperature dependence 
of the local bond susceptibility shows the existence of degenerate or almost 
degenerate states with different magnitude of their bond value $B_{nn}$. 
This doping range roughly corresponds to the pseudogap region of an $N_c=4$
cluster, see Fig.~\ref{fig:phasediagram}.  As shown  in Fig.~\ref{fig:chiaf} -c) 
the extrapolation of our data indicate that $\chi_0$ is diverging when  
$T \rightarrow 0$.  Since the lowest temperature we can reach is $\approx 0.01 t$ 
the  apparent divergent $\chi_0$ implies, if not degenerate states, at least bond 
excitations with an energy much smaller than this scale.

\begin{figure}[t]
\begin{center}
\includegraphics*[width=3.3in]{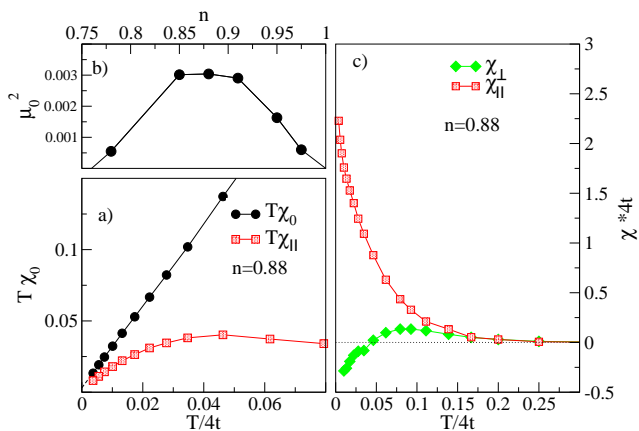}
\caption{(color online) a) $T\chi_0(T)$ show a linear behavior and extrapolate to
a finite value $\mu^2_0$ at $T=0$. b)$\mu^2_0$ versus filling $n$.
c) $\chi_{\parallel}(T)$ increases strongly at low $T$,
presumably diverging. However the divergence is weaker than 
$\sim \frac{1}{T}$ as $T \chi_{\parallel}(T)$ in b) shows.
 $\chi_{\perp}(T)$  is small and negative at low $T$. }
\label{fig:muloc}
\end{center}
\end{figure}

In the pseudogap region (i.e.,  $1>n>\sim 0.78$) $\chi_0$ diverges as 
$\sim \frac{1}{T}$. This can be seen in Fig.~\ref{fig:muloc} -a) where 
we show (black line)  $T\chi_0$ versus $T$ at filling $n=0.88$. $T\chi_0$ 
displays a linear behavior and, at $T=0$, extrapolates to a finite value, 
albeit small, $\mu^2_0$.  The $\sim \frac{1}{T}$ dependence of susceptibility
is consistent with scenarios which assume two degenerate configurations, 
$1$ and $2$, with different  bond values such that $2 \mu_0=B_{nn}(1)-B_{nn}(2)$.
It is instructive to draw an analogy with local spin susceptibility of  a system with 
independent local moments.  A free moment is a doubly degenerate problem 
where the spin can be aligned parallel or antiparallel to a particular direction. 
A perturbing magnetic field  lifts the degeneracy of the two configuration by
an amount proportional to the moment and the magnetic field.   Similarly, in 
our system the perturbing Hamiltonian acting on the bond $\langle ij \rangle$, 
$H^{ext}=h B(i,j)$, splits the two configurations with an amount proportional 
to the field $h$ and $B(i,j)(1)-B(i,j)(2)$.  Thus, the $\sim \frac{1}{T}$ dependence 
of the local bond susceptibility suggests  the existence  of two degenerate 
states with different bond magnitude. Of course other scenarios compatible with
a $\sim \frac{1}{T}$ like susceptibility cannot be excluded.

The bond correlations are strongest around $n \approx 0.88$ and weak at 
small and large doping.  This can be seen both by inspecting $\chi_0(T)$ 
in Fig.~\ref{fig:chiaf} -c) and by looking at the bond moment $\mu_0^2$ 
versus filling in Fig.~\ref{fig:muloc} -b).   At half filling, the bond moment 
extrapolates to zero, since the numerical data does not show evidence of 
divergent susceptibilities in the undoped system.  At finite doping  $\mu_0^2$ 
increases with increasing doping displaying a maximum at $n \approx 0.88$.  
$\mu_0^2(n)$ decreases with further doping until it vanishes at $n \approx 0.78$.

At low temperatures we find a strong positive correlation between nearest 
neighbor parallel bonds and a small negative correlation between nearest 
neighbor perpendicular bonds. This is shown in Fig.~\ref{fig:muloc} -c.
$\chi_{\parallel}$ is increasing strongly with lowering $T$, 
the numerical data indicating even a possible divergence when 
$T \rightarrow 0$, although weaker than $\sim \frac{1}{T}$ characteristic 
to local bond fluctuations (see Fig.~\ref{fig:muloc} -a).  The large 
value of $\chi_{\parallel}$ shows that increasing or reducing the 
antiferromagnetism on a bond implies a similar effect on the nearest 
neighbor parallel bond.  Whereas the correlation between nearest neighbor 
perpendicular bond fluctuations $\chi_{\perp}$  is much smaller 
and negative.


\begin{figure}[t]
\begin{center}
\includegraphics*[width=3.3in]{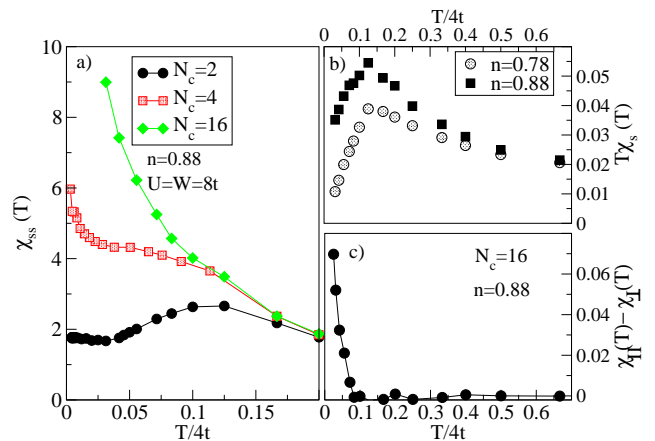}
\caption{(color online) Susceptibility $\chi_{s}(T)$ at $n=0.88$ for different cluster sizes.
The divergence at $T=0$ is more pronounced for the large $N_c=16$ cluster.
No divergence is present for $N_c=2$.  b) $T\chi_{s}(T)$ when $N_c = 16$
for two different fillings. c) $\chi_{\parallel}$ -  $\chi_{\perp}$ at $n=0.88$ for $N_c=16$. }
\label{fig:RVB_vsNC}
\end{center}
\end{figure}

Larger cluster calculations are limited to finite temperatures due to 
the minus-sign problem present in the Hirsch-Fye algorithm.
For example, for the values of $U/W$ used here, when $N_c=16$ 
the minus sign limits Hirsch-Fye QMC calculations to $4t\beta  \le 36$ for 
fillings in the pseudogap region. Down to this temperature the local bond 
correlation $\chi_{0}(T)$ looks similar to the one calculated for $N_c=4$, 
but this temperature is too large for a reliable extrapolation to $T=0$. 
However,  by increasing the cluster size and thus 
incorporating  spatial correlations at larger length scale one expects a 
decrease of  $\mu_0$ or even the disappearance of the divergent behavior 
due to the settlement  of bond order. This is similar to the disappearance of 
the $T=0$ divergence in the local spin susceptibility when going from an
atom  to finite cluster due to the settlement of short range AF order.

In order to investigate critical fluctuations as a function of the cluster size we 
calculate the cluster bond correlation  $\chi_{s}(T)$ defined in Eq.~\ref{eq:chis}.  
While the zero temperature divergence of  $\chi_{0}(T)$ is suppressed  in larger 
clusters due to correlations between the bonds, $\chi_{s}(T)$ would still be 
divergent provided that short ranged order between the bonds emerges.
The plots in  Fig.~\ref{fig:RVB_vsNC} -a)  of  $\chi_{s}(T)$ for $N_c=2$, 
$N_c=4$ and $N_c=16$ at $n=0.88$  indicate  that unquenched  bond fluctuations  
persists with increasing cluster size.  For $N_c=2$ (black), $\chi_{s}(T)$, which 
in this case coincides with the local bond $\chi_{0}(T)$, does not show  
divergent behavior at zero temperature.  $\chi_{s}(T)$ for $N_c=4$  
diverges at $T=0$ since it contains the divergent $\chi_{0}(T)$ term.  For 
$N_c=16$  and at the accessible temperatures,  $\chi_{s}(T)$ increases 
more strongly with decreasing temperature than for the $N_c=4$ case, thus 
showing an even more divergent behavior. Of course, the large value of $\chi_{s}$ 
for $N_c=16$ show the importance  of spatial correlations between bond 
fluctuations. As for the $N_c=4$ cluster, we find that for $N_c=16$ the
correlations between nearest neighbor parallel bonds is more important than 
the correlations between nearest neighbor perpendicular bonds,
as the difference between $\chi_{\parallel}$ and  $\chi_{\perp}$ 
plotted in Fig.\ref{fig:RVB_vsNC}-c  shows.

Notice that  even for the $N_c=16$  cluster  the divergent  behavior of  
$\chi_{s}(T)$  ceases when $n< \approx 0.78$ as can be seen in  of 
Fig.~\ref{fig:RVB_vsNC} -b,  indicating that $n \approx 0.78$ is a QCP.
The smaller (larger) doping region would presumably correspond to a state 
with (without) bond order when $N_c \rightarrow  \infty$.

\begin{figure}[t]
\begin{center}
\includegraphics*[width=3.3in]{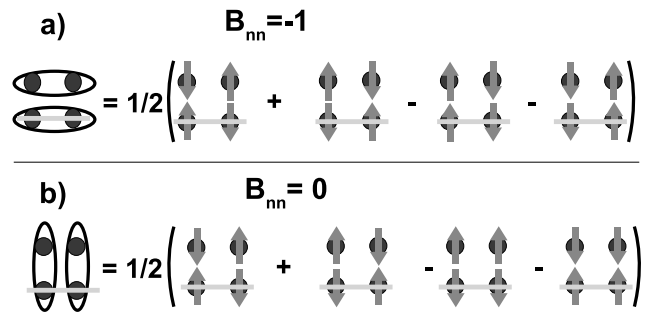}
\caption{a) Configuration with  bond singlets along $x$ direction. The marked bond is a superposition of 
states with AF aligned spins. b)  Configuration with  bond singlets along $y$ direction. The marked bond 
is a superposition  of two states with AF aligned spins and of two states with FM aligned spins. 
The bond operator, Eq.~\ref{eq:Bi}, measured 
on the bond along $x$ direction (marked bond) takes the value $B_{nn}=-1$ ($B_{nn}=0$) 
for configuration -a (-b). If a) and  b) are degenerate, the local bond susceptibility will diverge $\propto \frac{1}{T}$ 
when $T\rightarrow 0$.
 }
\label{fig:singlets}
\end{center}
\end{figure}

{\em{Discussion}}
Without dismissing other possibilities, we note that a scenario where the 
system forms adjacent parallel bond singlets fits very well with our 
results. For instance the divergence of local bond susceptibility $\chi_{0}$ 
requires two degenerate states with different $B_{nn}$.  Suppose we measure 
$\chi_{0}$ on a bond along $x$ direction.  A configuration with adjacent 
parallel  bond singlets along $x$, such as one in Fig.~\ref{fig:singlets} a),
has a $B_{nn}=-1$, while a configuration with adjacent parallel  bond singlets 
along $y$, such as one in Fig.~\ref{fig:singlets} b), has a $B_{nn}=0$. If these 
two configurations are degenerate they will yield a divergent $\chi_{0}$. 
Moreover  the correlation between parallel bond excitations will be also 
divergent and positive, while the correlation between nearest neighbor 
perpendicular bonds will be negative,  resembling our findings on the 
$2 \times 2$ cluster.

{\em{Conclusions.}}  
The behavior of  bond  susceptibility in  the PG region of the 2D Hubbard 
model calculated with DCA shows evidence of  quantum criticality and implies 
settlement of  bond order.  Thus, in the the $2\times2$ cluster we find 
divergent local bond susceptibility at $T=0$, which implies ordering in 
the imaginary time due to  the existence of  degenerate bond spin  excitations.
We find a strong correlation between  excitations on parallel neighboring 
bonds, which suggests that the pseudogap region is characterized by bond 
singlet dimerization.  We argue that the existence of unquenched local 
zero energy fluctuations for small $N_c$ implies long range order in the 
limit $N_c \rightarrow \infty$ or the intervention of competing phase 
transition.  The suppression of  divergence for $n< \approx 0.78$ 
implies that $n \approx 0.78$ is a QCP which separates the unconventional
pseudogap region characterized by dimers from a conventional Fermi liquid.


\acknowledgments  This research was supported by NSF DMR-0312680, 
DMR-0706379 and CMSN DOE DE-FG02-04ER46129.  Supercomputer support was 
provided by the Texas Advanced Computing Center.  We would like to 
thank
J.C.\ Davis, 
P.\ Hirschfeld, 
M.\ Ma, 
M.\ Norman,
G.\ Sawatzky, 
and 
N.S.\ Vidhyadhiraja
for stimulating conversations, and all of the organizers of the Sanibel 
Symposium.

\end{document}